\documentclass[aps,preprint,amsmath,amssymb,amsfonts,nofootinbib]{revtex4}
\usepackage{epsfig}
\usepackage{graphicx}
\usepackage{dcolumn}
\usepackage{bm}
\usepackage{amsthm}
\usepackage{amsmath}
\usepackage{color}

\newcommand{\beq}{\begin{equation}}
\newcommand{\eeq}{\end{equation}}
\newcommand{\bea}{\begin{eqnarray}}
\newcommand{\eea}{\end{eqnarray}}

\begin{document}

\vskip 1cm

\title{Holographic Vorticity in the Fluid/Gravity Correspondence}
\author{Christopher Eling$^1$}
\author{Yaron Oz$^2$}

\affiliation{$^1$ Max Planck Institute for Gravitational Physics, Albert Einstein Institute, Potsdam 14476, Germany}
\affiliation{$^2$ Raymond and Beverly Sackler School of Physics and Astronomy, Tel Aviv University, Tel Aviv 69978, Israel}

\date{\today}

\begin{abstract}

The vorticity statistics characterizes both the direct and the inverse turbulent cascades of two-dimensional fluid flows.
The fluid/gravity correspondence relates fluid flows to black brane dynamics.
We construct the holographic vorticity for relativistic and non-relativistic fluids in terms of the gravitational black brane data, and relate
it to the horizon vorticity expressed as a Weyl scalar.
We discuss the statistical scaling structure of the horizon geometry.

\end{abstract}


\maketitle

\tableofcontents

\newpage

\section{Introduction and Summary}

Ideal fluid flows in $2+1$ space-time dimensions have two positive quadratic conserved quantities, the energy and the enstrophy (vorticity squared).
As such, when exciting the fluid one generates two turbulent cascades, the direct and the inverse. In the direct cascade the enstrophy flows to length scales smaller than
the excitation scale, while in the inverse cascade energy flows to length scales larger than the excitation scale.
The vorticity statistics and its scaling properties characterizes both the direct and the inverse turbulent cascades of two-dimensional fluid flows
\cite{tab}.

In the fluid/gravity correspondence framework, fluid flows are related to black brane dynamics.
The aim of this paper is to construct the holographic relativistic and non-relativistic vorticity in terms of the gravitational black brane data. We will then outline
the expected statistical scaling structure of the geometry. In the following we summarize the results.

The ideal stress-energy tensor of relativistic CFT hydrodynamics in flat $2+1$ space-time dimensions reads (up to an overall constant)
\begin{equation}
T^{\mu\nu} = T^3\left( \eta^{\mu\nu} + 3 u^{\mu}u^{\nu} \right) \ ,
\end{equation}
where $T$ is the temperature, and $u^{\mu}$ is the 3-velocity vector satisfying $u_{\mu}u^{\mu}=-1$.
The three independent degrees of freedom of the hydrodynamics are encoded in the covector $c_{\mu}\sim Tu_{\mu}$, which will appear naturally in the
gravitational description.
The hydrodynamics equations $\partial_{\mu}T^{\mu\nu}=0$ can be written in a projected form as
\begin{align}
P_{\nu\sigma}\partial_{\mu}T^{\mu\sigma} = \Omega_{\mu \nu} u^\mu = 0,~~~~~~u_{\nu}\partial_{\mu}T^{\mu\nu} = \partial_\mu s^\mu =  0 \ ,
\label{hydroeq}
\end{align}
where $P_{\mu\nu} = \eta_{\mu\nu} + u_{\mu}u_{\nu}$, $s^{\mu} = T^2u^\mu$ is the entropy current,  and $\Omega_{\mu\nu}$ is a two-form that is used to define the
relativistic enstrophy
\begin{align}
\Omega_{\mu \nu} = \partial_{[\mu} (T u_{\nu]})  \ . \label{vorticity}
\end{align}

The first equation in (\ref{hydroeq}) is the relativistic Euler equation, and the second is the conservation of the entropy current.
We can write the two-form (\ref{vorticity}) as
\begin{align}
\Omega_{\mu \nu} = \xi \epsilon_{\mu\nu\lambda}u^{\lambda} \ , \label{xi}
\end{align}
for some real scalar function $\xi$.

The fluid dynamics is encoded in the dynamics of an event horizon in the gravitational dual description in one higher space dimension. The fluid temperature is mapped
to the horizon surface gravity, while the fluid velocity corresponds to the horizon normal.
One can define a complex function in terms of the bulk Weyl tensor $C_{ABCD}$,
\begin{equation}
\Psi_2 = C_{ABCD} \ell^A m^B \bar{m}^C n^D \ ,
\end{equation}
where $(\ell^A, m^B, \bar{m}^C, n^D)$ is a null tetrad basis in the Newman-Penrose formalism. $\Psi_2$ is one of the five complex Weyl  scalars that represent
the ten real variables of the Weyl tensor.
Its imaginary part is the horizon vorticity and is a measure for the strength of the precession of an infalling observer.
The real part, called tendicity,  is a measure the strength of the tidal stretching of an infalling observer.
We will see that the function $\xi$ is given by the imaginary part of $\Psi_2$,
$\xi = {\rm Im}\Psi_2$, while up to the cosmological constant $\Lambda$, the real part of $\Psi_2$ is
proportional to the divergence of the velocity 3-vector ${\rm Re}\Psi_2-\Lambda/6 \sim \partial_\mu u^\mu $.

When studying non-relativistic flows, the hydrodynamic equations (\ref{hydroeq}) become the incompressible Euler
equations. Since the velocity vector is divergence free $\partial_iv^i = 0$, it is completely determined by its curl, that is by the vorticity scalar
$\omega= \epsilon_{ij}\partial^iv^j$ satisfying the Euler equation $\frac{d\omega}{dt}=0$.
Indeed, we will show that $\omega \sim {\rm Im}\Psi_2$ and that it determines the horizon geometry uniquely.
The real part of $\Psi_2$ is just a constant $\Lambda/6$ in this case.

The paper is organized as follows.
In the section 2 we briefly review the structure of the horizon dynamics and derive the relativistic and non-relativistic Euler equations
from the horizon geometry. In section 3 we construct the holographic vorticity in both the relativistic and non-relativistic cases.
We show that the horizon geometry is uniquely determined in the non-relativistic case by one real scalar function corresponding to
the fluid vorticity. In section 4 we discuss the expected scaling structure of the horizon geometry in the turbulent regime.

\section{The Horizon Geometry}

\subsection{Review of null surfaces dynamics}

In the following we will briefly review some of the analysis  of \cite{Eling:2010hu} on the dynamics of null surfaces, which will be relevant
for us.
Denote the four coordinates of the bulk space-time by $x^A = (r, x^\mu),\mu=0,1,2$. $x^\mu$ are coordinates on the horizon $\mathcal{H}$, and $r$ is a transverse coordinate to it.
One can choose $r = 0$ as the location of the horizon.
We denote the null normal to the horizon by $\ell^A $. In fact it is both normal and tangent to the horizon and tangent to its null generators. In components
$\ell^A = (0, \ell^\mu)$.

The pullback of the bulk metric $g_{AB}$ into $\mathcal{H}$ is the degenerate horizon metric $\gamma_{\mu\nu}$. Its null directions are the generating light-rays of $\mathcal{H}$, i.e. $\gamma_{\mu\nu}\ell^\nu = 0$.
The Lie derivative of $\gamma_{\mu\nu}$ along $\ell^\mu$ is the second fundamental form
\begin{align}
	\theta_{\mu\nu} = \frac{1}{2}\mathcal{L}_{\ell}\gamma_{\mu\nu} \label{eq:theta_def} \  .
\end{align}
We can decompose $\theta_{\mu\nu}$ into the shear tensor $\sigma^{(H)}_{\mu\nu}$ and the expansion  $\theta$:
\begin{align}
	\theta_{\mu\nu} = \sigma^{(H)}_{\mu\nu} + \frac{1}{2}\theta\gamma_{\mu\nu}  \label{eq:theta_decompose} \ .
\end{align}
In fact, $\theta_{\mu\nu}=0$ at the ideal fluid order that we will be working with.

Since $\gamma_{\mu\nu}$ is degenerate, one cannot use it to define an intrinsic connection on the null horizon, as could be done for spacelike or timelike hypersurfaces. The bulk spacetime connection induces a notion of parallel transport in $\mathcal{H}$, but only along its null generators. This structure is not fully captured by $\gamma_{\mu\nu}$; instead, it is encoded by the extrinsic curvature $\Theta_\mu{}^\nu$, which is the horizon restriction of $\nabla_A\ell^B$
\begin{align}
	\Theta_\mu{}^\nu = \nabla_\mu\ell^\nu \ .
\end{align}

For a non-null hypersurface, the extrinsic curvature at a point is independent of the induced metric. For null hypersurfaces, this is not so. Indeed, lowering an index of $\Theta_\mu{}^\nu$ with $\gamma_{\mu\nu}$, we get the shear/expansion tensor $\theta_{\mu\nu}$
\begin{align}
	\Theta_\mu{}^\rho\gamma_{\rho\nu} = \theta_{\mu\nu}  \ . \label{eq:Theta_theta}
\end{align}
This expresses the compatibility of the parallel transport defined by $\Theta_\mu{}^\nu$ with the horizon metric $\gamma_{\mu\nu}$. Contracting $\Theta_\mu{}^\nu$ with $\ell^\mu$ yields the surface gravity $\kappa$, which measures the non-affinity of $\ell^\mu$:
\begin{align}
	\Theta_\mu{}^\nu\ell^\mu = \kappa\ell^\nu. \label{eq:Theta_kappa}
\end{align}
At the ideal fluid  order $\Theta_\mu{}^\nu$ can be written as
\begin{align}
	\Theta_\mu{}^\nu =  c_\mu\ell^\nu,~~~\quad c_\mu\ell^\mu = \kappa  \label{eq:Theta} \ .
\end{align}
In the literature on null horizon dynamics, $c_\mu$ is known as the ``rotation one-form" of the horizon \cite{Gourgoulhon:2005ng}. As we will see,
 $c_\mu$ encodes in its three components all the fluid data, that is the temperature and
the velocity 3-vector. The null Gauss-Codazzi equations that will give the ideal order Navier-Stokes equations  can be written as \cite{Eling:2010hu}
\begin{align}
	c_\mu \partial_\nu S^\nu + 2S^\nu\partial_{[\nu}c_{\mu]}  = 0 \ ,\label{eq:GC_basic}
\end{align}
where we defined the area entropy current $S^\mu = v \ell^\mu$, and
$v$ is a scalar density equal to the horizon area density. Note that the entropy current denoted here by $S^\mu$, corresponds under the map between
gravity and the hydrodynamics variables to $s^{\mu}$ in (\ref{hydroeq}).

\subsection{The fluid/gravity correspondence}

\subsubsection{Relativistic hydrodynamics}

Consider the boosted black brane metric in $3+1$ dimensions %
\begin{align}
ds^2 = -2 u_\mu dx^\mu dr - F(r) u_\mu u_\nu dx^\mu dx^\nu + G(r) P_{\mu \nu} dx^\mu dx^\nu, \label{eq:BBmetric}
\end{align}
where $F(0)=0$ is the horizon. The horizon quantities defined in the previous section can be expressed in terms of the Bekenstein-Hawking entropy density $s= v/4= G(0)/4$ and the Hawking temperature $T= \kappa/2\pi=-F'(0)/2$
\begin{align}
 S^{\mu} &= 4 s u^\mu; & \Theta_{\mu}{}^{\nu} &= -2\pi T u_\mu u^\nu; & \gamma_{\mu\nu} &= (4s) P_{\mu\nu}; & c_\mu &= -2\pi T u_\mu \ . \label{eq:Zeroth}
\end{align}
A particular example in this class of metrics is the Anti de-Sitter (AdS) black brane, which satisfies
the vacuum Einstein equations with negative cosmological constant (we use units where $\Lambda=-3$)

\begin{align}
R_{AB}+3 g_{AB} = 0 \ . \label{equations}
\end{align}
In this case $F= (r+R)^2 f(r+R)$, where $f = 1-(\frac{R}{r})^3$, and $G(r) = (r+R)^2$. Thus, $s= R^{2}/4$ and the Hawking temperature $T= \kappa/2\pi=(4/3\pi) R$.

Consider  \eqref{eq:BBmetric}, but with $(u^\mu,T)$ slowly varying functions of $x^\mu$ rather than being  constants. This is no longer an exact  solution of (\ref{equations}),
but rather an approximate solution at zeroth order in derivatives \cite{Bhattacharyya:2008jc}. Plugging in the zeroth order values (\ref{eq:Zeroth}) in Gauss-Codazzi equations \eqref{eq:GC_basic} and projecting along $u^\mu$, we find the conservation of the entropy current (\ref{hydroeq})
\begin{align}
\partial_\mu (s u^\mu) = \theta = 0 \label{entropyconsv} \ .
\end{align}
This means that the horizon at $r=0$ in the geometry (\ref{eq:BBmetric}) is non-expanding at this order in derivatives.
Projecting transverse to $u^\mu$ we get the ideal relativistic Euler equation (\ref{hydroeq}).
Using the definition (\ref{vorticity}) and the zeroth order expressions (\ref{eq:Zeroth}) we see
that
\begin{equation}
\Omega_{\mu \nu}  \sim \partial_{[\mu} c_{\nu]} \ .
\end{equation}
One can construct a conserved local enstrophy current \cite{Carrasco:2012nf}
\begin{align}
J^\mu = T^{-2} \Omega_{\alpha \beta} \Omega^{\alpha \beta} u^\mu \ ,
\end{align}
and correspondingly a conserved relativistic enstrophy
\begin{align}
{\cal Z} = \int d^2 x T^{-2} \Omega_{\alpha \beta} \Omega^{\alpha \beta} u^0   \label{Zrel} \ .
\end{align}

\subsubsection{Non-relativistic hydrodynamics}

The Navier-Stokes equations for non-relativistic fluid flows $v\ll c$ is obtained by introducing a small scaling parameter $\varepsilon$ such that
$\partial_i \sim \varepsilon, \partial_t \sim \varepsilon^2, v^i \sim \varepsilon$ and $T= T_0(1+ \varepsilon^2 p(x))$  for constant $T_0$
\cite{Fouxon:2008tb,Bhattacharyya:2008kq}.
The relativistic ideal CFT hydrodynamics equations (\ref{hydroeq}) become the non-relativistic incompressible Euler equations
\begin{align}
\partial_t v_i + v^j \partial_j v_i + \partial_i p = 0, ~~~~ \partial_i v^i = 0  \ . \label{NS}
\end{align}
where $v^i,i=1,2$ is the velocity vector field and $p$ is the fluid pressure.
Taking a divergence on both sides of (\ref{NS}) we see that the pressure is not an independent variable
$\nabla^2 p = -\partial_i v^j \partial_j v^i$. Also, since the vector field is divergence free, one needs to know only its curl in order to determine it.

In terms of the vorticity of the flow $\omega$ we have
\begin{equation}
\partial_t \omega + v^i\partial_i \omega = 0 \ .
\label{NSw}
\end{equation}
We see that in the absence
of friction and an external force there are two conserved quantities, the energy of the flow
\begin{equation}
{\cal E} = \int d^2 x \frac{v^2}{2} \ ,
\end{equation}
and the enstrophy of the flow
\begin{equation}
{\cal Z} = \int d^2 x \frac{\omega^2}{2} \ . \label{enstrophyNR}
\end{equation}
In the direct cascade the enstrophy is transfered to small length scales (compared to the force scale), while in
the inverse cascade the energy is transferred to large scales.
The relativistic two-form (\ref{vorticity}) reduces to the non-relativistic vorticity two-form
\begin{align}
\Omega_{\mu \nu} \rightarrow T_0 \partial_{[i} v_{j]} = T_0 \omega_{ij} \  , \label{reduce}
\end{align}
and the conserved relativistic entropy (\ref{Zrel}) becomes the conserved non-relativistic one (\ref{enstrophyNR}).

While we will be interested mostly in ideal hydrodynamics, note that if we include the first order corrections
to the gravitational description we will get a viscous term $\nu\partial_{jj} v^i$ on the RHS of the Euler equation (\ref{NS})
with kinematic viscosity  \cite{Eling:2010hu}
\begin{equation}
\nu = \frac{1}{4\pi T_0} \ . \label{kinematic}
\end{equation}
In deriving (\ref{kinematic}), one uses the ratio of the shear viscosity to entropy density in Einstein gravity $\frac{\eta}{s} = \frac{1}{4\pi}$  \cite{Kovtun:2003wp}.

\section{Vorticity and a Geometric Charaterization of the Horizon}

Consider the identity formula defining the Riemann tensor,
\begin{align}
(\nabla_A \nabla_B - \nabla_B \nabla_A)\ell^C = R_{ABD}{}^C \ell^D \ ,
\end{align}
where $\ell^A$ is the horizon normal. We now take the pullback to the horizon (in the following all equalities will hold at the horizon) and use (\ref{eq:Zeroth}),
which is valid for the fluid background at lowest order in derivatives. We get
\begin{align}
2 \nabla_{[\mu} c_{\nu]} \ell^C = - R_{\mu \nu D}{}^C \ell^D = -  C_{\mu \nu D}{}^C \ell^D \label{Weyl1} \ ,
\end{align}
where $C_{ABCD}$ is the Weyl tensor. The last equality holds because we imposed $R_{\mu B} \ell^B=0$. This reduces to
\begin{align}
2 \nabla_{[\mu} c_{\nu]} u^\rho = - C^{(1)}_{\mu \nu \lambda}{}^\rho u^\lambda,
\end{align}
where the superscript refers to the derivative order of the Weyl tensor.

We introduce now a null tetrad basis $(\ell^A, n_A, m^A, \bar{m}^A)$. This basis is defined in the following way: $n_A$ is a null co-vector satisfying $n^A \ell_A = -1$,
the $m^A$ and $\bar{m}^A$ are complex null vectors satisfying $m_A \ell^A = m_A n^A = \bar{m}_A n^A = \bar{m}_A \ell^A = 0$, and $m_A \bar{m}^A = 1$. If we contract (\ref{Weyl1}) with $n_A$, we find
\begin{align}
2 \nabla_{[\mu} c_{\nu]} =  C_{\mu \nu D}{}^C n_C \ell^D \label{Weyl2} \ .
\end{align}
In our fluid setup in the $(r,x^\mu)$ coordinates
\begin{align}
\ell_A = (1,0), \ell^A = (0,u^\mu); ~~~ n_A = (0, u_\mu), n^A = (1,0) \ ,
\end{align}
while the complex $m$ and $\bar{m}$ always have $\mu$ components and are orthogonal to $u^\mu$. This implies that
\begin{align}
2 \nabla_{[\mu} c_{\nu]} = C^{(1)}_{\mu \nu \lambda r} u^\lambda \  ,
\end{align}
which is just an identity that we have checked holds for the zeroth order metric (\ref{eq:BBmetric}).

Contracting (\ref{Weyl1}) with any vector $v^A$, such that $v_A \ell^A = 0$ yields
\begin{align}
C_{\mu \nu D}{}^C v_C \ell^D = 0 \ ,
\end{align}
for a non-expanding horizon.
Thus, expanding out $C_{ABCD} \ell^C n^D$ in terms of the various Weyl scalars and pulling back to the horizon leads to the following result \cite{Ashtekar:2000hw}
\begin{align}
C^{(1)}_{\mu \nu \lambda r} u^\lambda = 4i {\rm Im}[\Psi_2] m_{[\mu} \bar{m}_{\nu]} \ .
\end{align}
This  implies
\begin{align}
\nabla_{[\mu} c_{\nu]} = 2 i {\rm Im}[\Psi_2] m_{[\mu} \bar{m}_{\nu]}.
\end{align}
In terms of the fluid variables on the horizon, we can use
\begin{align}
2 i m_{[\mu} \bar{m}_{\nu]} = u^\lambda \epsilon_{\lambda \mu \nu},
\end{align}
and relate $\xi$ (\ref{xi}) to the the Weyl scalar
\begin{align}
\xi = {\rm Im} \Psi_2  \ .
\end{align}

In the non-relativistic limit (\ref{reduce})
\begin{align}
\omega  = \frac{1}{2T_0} {\rm Im}\Psi_2 \ . \label{relation}
\end{align}
Thus, the imaginary part of $\Psi_2$ gives a complete characterization of the fluid vorticity.
Furthermore, although the NP formalism contains some ``gauge" arbitrariness due to the freedom to select any null tetrad, in the case of a non-expanding horizon
$\Psi_2$ is gauge invariant, i.e. is independent of this choice \cite{Ashtekar:2000hw}.

The second variable characterizing the horizon geometry is the intrinsic scalar curvature on the cross-sections, $\tilde{R}$. Following \cite{Ashtekar:2004gp} one can define the complex function
\begin{align}
\Phi_H = \frac{1}{4} \tilde{R} - i {\rm Im} \Psi_2 \  .
\end{align}
In terms of $\Psi_2$ and the bulk curvature this quantity can be re-expressed as
\begin{align}
\Phi_H = -\Psi_2 + \frac{1}{4} R_{AB} q^{AB} - \frac{1}{12} R,
\end{align}
where $q^{AB}$ is the projector onto the cross-sections. For our generic boosted black brane metric (\ref{eq:BBmetric}),
\begin{align}
q^{AB} = \frac{1}{G} P^{\mu \nu}.
\end{align}
Thus, ${\rm Re} \Phi_H = -{\rm Re} \Psi_2 + (1/4) G^{-1} P^{\mu \nu} R_{\mu \nu} - (1/12) R$. Note that in the vacuum setting, when the cosmological constant (and any matter fields) are not present, the additional curvature terms vanish and ${\rm Re} \Phi_H = -{\rm Re} \Psi_2$.

We want to calculate this quantity to first order in derivatives. We use the formula \cite{Ashtekar:2000hw}
\begin{align}
{\rm Re} \Psi_2 = \frac{1}{2} C_{ABCD} \ell^A n^B \ell^C n^D.
\end{align}
At zeroth order ${\rm Re} \Psi_2 = -\frac{1}{2}$ and the full expression ${\rm Re} \Phi_H$ vanishes. This is as expected, since the curvature of the planar horizon cross-sections should vanish. At first order, the formula leads to
\begin{align}
{\rm Re} \Psi^{(1)}_2 = \frac{1}{2} C^{(1)}_{\mu r \nu r} u^\mu u^\nu,
\end{align}
expressing the real part of this Weyl component in terms of the fluid variables. Putting everything together and imposing the ideal order equation
\begin{align}
\frac{u^\lambda \partial_\lambda G}{G} = -\partial_\lambda u^\lambda,
\end{align}
we find
\begin{align}
{\rm Re} \Phi^{(1)}_H = \frac{1}{2} \frac{u^\lambda \partial_\lambda G'}{G} - \frac{1}{8} \frac{G' \partial_\lambda u^\lambda}{G} \ ,
\end{align}
where prime denotes derivatives with respect to the radial coordinate $r$.
Generically,
\begin{align}
\frac{u^\lambda \partial_\lambda G}{G} \sim u^\lambda \partial_\lambda T \\
\frac{G'}{G} \sim \frac{1}{T} \ .
\end{align}

Using (\ref{hydroeq}) we get
\begin{align}
{\rm Re} \Phi^{(1)}_H \sim \frac{\partial_\lambda u^\lambda}{T} \ ,
\end{align}
In the specific case of the AdS black brane, the proportionality constant is $-9/16\pi$. Note that ${\rm Re} \Psi^{(1)}_2$, a gauge invariant quantity in its own right, is also by itself proportional to $\frac{\partial_\lambda u^\lambda}{T}$.  Taking the non-relativistic limit we have
\begin{align}
{\rm Re} \Phi^{(1)}_H \sim \partial_i v^i \ .
\end{align}
Thus, in the gravitational description of the two-dimensional non-relativistic incompressible fluid the real function ${\rm Im}\Psi_2$ characterizes
completely the horizon geometry. This is in parallel to the vorticity that characterizes fully the dual fluid flow.

\section{Geometrical Scalings from Turbulence Cascades}

Consider a random external force that excites the two-dimensional fluid at a scale $L$. In general, the statistics of the flow velocity and vorticity that in generated by the excitation
is not calculable due to the non-linear nature of the fluid equations.
An important question is, however, what are the universal characteristics of the flow statistics.
One distinguishes between a direct cascade and an inverse cascade of the fluid flow.
In the direct cascade the excitation of the fluid is at a large length scales, the dissipation is at small length scales, and
the enstrophy is transferred throughout the scales.
It is established numerically and experimentally that scale invariance is broken in the direct cascade and the correlations
functions of the vorticity are logarithmic \cite{tab}
\begin{equation}
\langle \omega^n(\vec{r},t)\omega^n(0,t)\rangle \sim \left[D \ln \left(\frac{L}{r}\right)\right]^{\frac{2n}{3}}  \  , \label{direct}
\end{equation}
where $D= \langle \nu(\nabla\omega)^2\rangle$ is the mean enstrophy dissipation rate.
In particular the energy spectrum is
\begin{equation}
E(k) \sim D^{\frac{2}{3}} k^{-3} \ln^{-\frac{1}{3}}\left(k L\right) \ .
\end{equation}

Relation (\ref{relation}) together with (\ref{direct}) implies, that once the horizon geometry is excited and becomes a random surface, the high momentum structure of the correlation functions
of ${\rm Im}\Psi_2$ should such exhibit a logarithmic structure.

In the inverse cascade, turbulence happens at length scales exceeding the force scale and energy is transferred throughout the scales.
 Here there is numerical and experimental evidence that
turbulence exhibits scale invariant statistics  \cite{tab}
and perhaps also conformal invariance \cite{falkovich}.
The Kolomogorov-Kraichnan scaling for the velocity $v_r$ and the vorticity $\omega_r$  at scale $r$
\begin{equation}
v_r \sim r^{\frac{1}{3}},~~~~\omega_r \sim r^{-\frac{2}{3}} \ , \label{vw}
\end{equation}
holds in the inverse cascade.
In particular, the energy spectrum
$E(k)$ has the scaling
\begin{equation}
E(k) \sim k^{-\frac{5}{3}}  \ .
\label{Kolmogorov}
\end{equation}
This scaling has been argued to be observed
in gravity \cite{Adams:2013vsa},   in the case of decaying turbulence.

Consider  an isoline of zero vorticity, that is the outer boundaries of a vorticity cluster of radius $l$.
The vorticity flux through the cluster scales like $l^{\frac{4}{3}}$. By relating this to the velocity circulation along the
boundary, it has been argued in \cite{falkovich}  that the perimeter scales like $l^{\frac{4}{3}}$. Thus,
the fractal dimension of the boundary curve is
\begin{equation}
d_{fractal} = \frac{4}{3} \ . \label{fractal}
\end{equation}
 This has been established numerically at the level of several percent accuracy in \cite{falkovich}  . Moreover, it has been shown
that the curve is a random SLE curve, signaling that scale invariance is in some sense enhanced to a conformal invariance.

Thus, the random horizon geometry at large length scales should exhibit
the scalings  (\ref{vw}) and  the fractal dimension (\ref{fractal}) of the isolines ${\rm Im}\Psi_2=0$.
It would be interesting to verify this structure numerically (see \cite{Owen:2010fa} for a numerical analysis of the horizon vorticity and tendicity
in black holes merger) .

 Note also that from the geometrical point of view we  can consider the statistics
of the same scalar function   ${\rm Im}\Psi_2$ for both the relativistic and the non-relativistic flows. It is therefore possible that the above
universal scale and conformal structures of the non-relativistic cascades may have an analog or perhaps even having its roots
in the turbulent CFT relativistic fluid flows.

\section*{Acknowledgements}

C.E. would like to thank Jose Luis Jaramillo for valuable discussions. This work is supported in part by the ISF center of excellence.


\end{document}